\documentclass[doublecol]{epl2} 

\usepackage{graphics}
\usepackage{epsfig}
\usepackage{amssymb}
\usepackage{subfigure}
\usepackage{pifont}

\pdfoutput=1

\title{Microrheology to probe non-local effects in dense granular flows}
\shorttitle{Microrheology to probe non-local effects in dense granular flows} 

\author{Mehdi Bouzid  \and Martin Trulsson \and Philippe Claudin \and Eric Cl\'ement \and Bruno Andreotti}
\shortauthor{M. Bouzid \etal}
\institute{Physique et M\'ecanique des Milieux H\'et\'erog\`enes, UMR 7636 ESPCI -- CNRS -- Univ.~Paris-Diderot -- Univ.~P.M.~Curie, 10 rue Vauquelin, 75005 Paris, France.}


\abstract{A granular material is observed to flow under the Coulomb yield criterion as soon as this criterion is satisfied in a remote but contiguous region of space. We investigate this non-local effect using discrete element simulations, in a geometry similar, in spirit, to the experiment of Reddy~\textit{et al}. \cite{RFP11}: a micro-rheometer is introduced to determine the influence of a distant shear band on the local rheological behaviour. The numerical simulations recover the dominant features of this experiment: the local shear rate is proportional to that in the shear band and decreases (roughly) exponentially with the distance to the yield conditions. The numerical results are in quantitative agreement with the predictions of the non-local rheology proposed by \cite{BTCCA13} and derived from a gradient expansion of the rheology $\mu[I]$. The consequences of these findings for the dynamical mechanisms controlling non-locality are finally discussed.}

\begin{document}

\maketitle

An ideal rheometer is supposed to measure the shear stress $\tau$ as a function of the shear rate $\dot \gamma$ in a steady homogeneous shear flow. Considering rigid grains confined under a pressure $P$, a linear shear flow profile allows to determine the function $\mu(I)$ relating the stress ratio $\tau/P$ to the inertial number $I=\dot \gamma d/\sqrt{P/\rho}$, for a density $\rho$ and a mean grain diameter $d$ \cite{GDRMidi,CEPRC05,JFP06}. The dimensionless number $I^2$ compares the inertial (Bernoulli-like) stress $\rho \dot \gamma^2 d^2$ to the confining pressure $P$. However, such an ideal rheometer does not provide complete information on the system: the \textit{local} constitutive law $\tau/P=\mu(I)$ is unable to explain crucial experimental features \cite{GDRMidi,A07,AFP13,DLDA06,P99}, hereafter referred to as \textit{non-local effects} and in particular, the possibility for the grains to flow even when the stress ratio $\tau/P$ is lower than the critical value $\mu(0) \equiv \mu_c$ \cite{KINN01,NZBWvH10,KK12,HK13,BTCCA13}. Non-locality in dense granular flow has first been related to the transmission of momentum at distances, through "force chains" \cite{A07}. Alternative mechanisms have recently been proposed, either based on localized plastic events \cite{PF09}, on dynamical heterogeneities \cite{DJvH} or on properties of the grain contact network \cite{LDW12,preprintArxiv}. 

\begin{figure}[t!]
\includegraphics{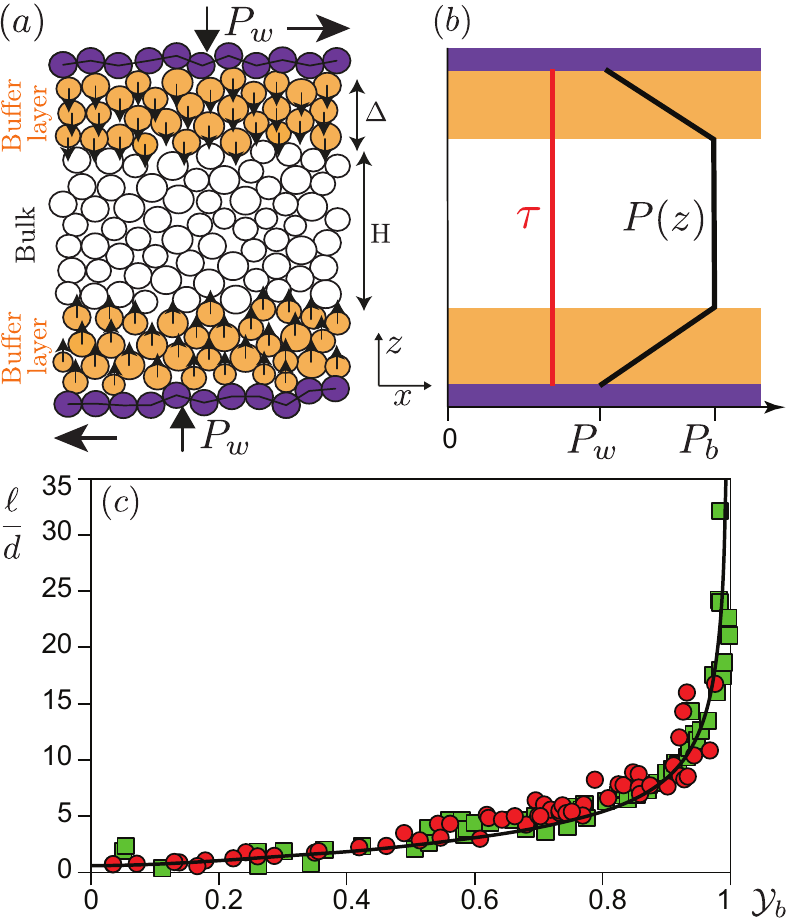}
\vspace{0 mm}
\caption{(a) Schematics of the numerical shear cell before the introduction of the micro-rheometer (see \cite{BTCCA13}). Bulk of the cell: $H \simeq 45 d$. Buffer layers: $\Delta \simeq 5d$. (b) Transverse shear stress profile $\tau$ (red line) pressure profile $P$ (black line), controlled by gravity-like bulk forces applied to the grains in the boundary layers located close to the walls. (c) Relaxation length $\ell$ of the grain velocity profile as a function of the bulk yield parameter $\mathcal{Y}_b = \tau/(\mu_c P_b)$. Data for $f=0$ (red circles) and  $f=0.4$ (green squares). Solid line: fit of the data providing a calibration of the non-local correction $\chi(\kappa)$ (Eq.~\ref{EqNonLocal}). Close to the yield conditions we have $\ell/d \sim \sqrt{\nu} / (1-|\mathcal Y_b|)^{1/2}$, with $\nu \simeq 8$ \cite{BTCCA13}.}
\vspace{0 mm}
\label{Fig1}
\end{figure}

In a recent experiment, Reddy~\textit{et al} \cite{RFP11} designed a micro-rheometer suited to determine the rheology in a region below the yield conditions ($\tau/P<\mu_c$). Under these conditions, a micro-probe is capable to flow under the remote influence of a shear zone forced by rigid boundaries. The central result is an exponential dependence of the local shear rate with the distance to the yield condition. Reddy~\textit{et al} have interpreted this dependence as a Boltzmann-like factor and suggested an analogy with the Eyring's transition state theory for the viscosity of liquids \cite{PF09}. In that case, mechanical fluctuations would play a role similar to the temperature in thermal systems. Assuming that the flow can be decomposed into elementary plastic rearrangements localized in space and well separated in time, each of these events would generate, at random, a new realization of the forces on the contact network, allowing for the formation of new weak zones preparing the next rearrangement\cite{TLB06,LC09,BCA09,LP09,ANBCC12}. In parallel to the ongoing discussion on the dynamical mechanisms responsible for non-locality, different models have been proposed to capture macroscopically the non-local effects \cite{at2001,volfson2003,aranson2008,AT06,KK12,BTCCA13}.

In this Letter, we challenge Reddy~\textit{et al}'s interpretation of the micro-rheological behaviour. By means of  discrete element numerical simulations, we reproduce a situation analogous to the experiment reported in \cite{RFP11}. Choosing a simpler and homogeneous configuration allows us to provide a clearer analysis of the phenomenon as well as  some analytical developments derived from a gradient expansion approach \cite{BTCCA13}.

\section{Numerical set-up}
We consider a two-dimensional system (Fig.~\ref{Fig1}a), similar to that used in~\cite{BTCCA13}, constituted of  $N \simeq 2\cdot 10^3$ circular particles of equal mass and of diameter randomly chosen in a flat distribution between $0.8\;d$ and $1.2\;d$. Such a polydispersity is an optimal compromise insuring that the sample remains homogeneous and does not crystalize. The grains are confined in a shear cell composed of two rough walls distant on the average by $\simeq 55 d$ and moving along the $x$-direction at opposite constant velocities. The walls are made of similar grains, glued together. Their positions are controlled to ensure a constant velocity and a constant normal stress $P_w$ at the walls -- the distance between the walls then fluctuates during the simulations (typically by a fraction of the grain diameter). Periodic boundary conditions are used along the $x$-direction. The particle (inertial) dynamics is integrated using the Verlet algorithm. Contact forces between particles are modeled as viscoelastic forces, with a Coulomb friction along the tangential direction \cite{CEPRC05,Cundall79,Luding06}. The normal spring constant $k_n$ is chosen sufficiently large (i.e. $k_n/P>10^{3}$) for the simulations to be in the rigid asymptotic regime where the results are insensitive to its value. The tangential spring constant is $k_t=0.5 k_n$. The Coulomb friction coefficient is chosen equal to $f=0.4$, but simulations have also been run with frictionless grains ($f=0$). Damping parameters are chosen such that the restitution coefficient is $e \simeq 0.9$. We have checked that our results are independent of the value of $e$. The reference local constitutive relation $\mu(I)$ for this system has been computed in \cite{BTCCA13}.

\begin{figure*}[t!]
\includegraphics{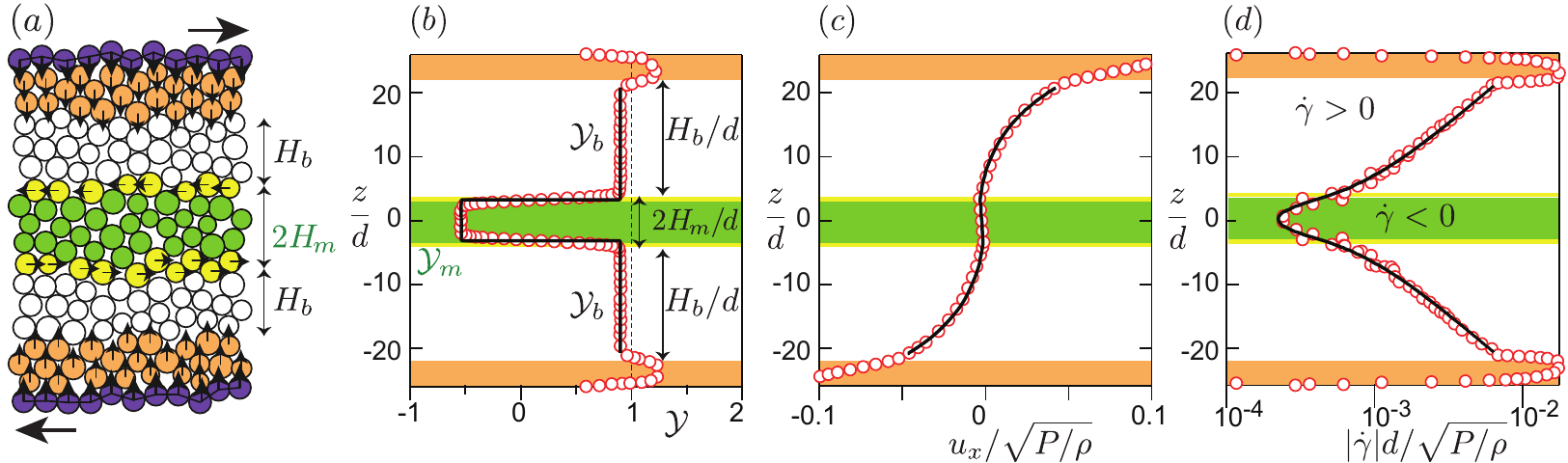}
\vspace{0 mm}
\caption{(a) Schematic of the numerical set-up with a micro-rheometer in the center (green zone in all panels). (b) Profile of the yield parameter $\mathcal{Y} = \tau/(\mu P)$, represented here for $\mathcal{Y}_b =0.9$ and $\mathcal{Y}_m =-0.5$. Corresponding velocity (c), shear rate (d) profiles. Symbols: numerical data. Solid lines: predictions of the non-local rheological model of \cite{BTCCA13}, assuming the continuity of $I$ and $dI/dz$ at $z=\pm H_m$, without any adjustable parameter.}
\vspace{0 mm}
\label{Fig2}
\end{figure*}

Following the numerical procedure of the present authors \cite{BTCCA13}, additional external forces are applied to the grains, whose strength and orientation depend on their transverse position $z$, as if there was an arbitrary gravitational field varying in space. These forces can either be along $z$, to vary the confining pressure $P$, or along $x$, to vary the shear stress $\tau$ (see schematics in Figs.~\ref{Fig1}a and \ref{Fig2}a). The profile of the yield parameter $\mathcal{Y}=\tau/(\mu_c P)$, which compares the stress ratio to the dynamical friction coefficient $\mu_c$, can therefore be imposed at will. The region under study is located in the bulk of the shear cell (denoted with a subscript $b$, as the pressure $P_b$ in Fig.~\ref{Fig1}b), of thickness $H$, where the stresses are homogeneous. The bulk is confined by two boundary layers, of thickness $\Delta$, located close to the walls, in which the normal stress $P$ is increased gradually from $P_w$ to $P_b$ thanks to these gravity-like vertical forces, while the shear stress $\tau$ is kept constant across the cell. As reported in \cite{BTCCA13}, the grains are observed to flow in the bulk even when the yield parameter $\mathcal{Y}_b=\tau/(\mu_c P_b)$ is smaller than $1$. In this case, the grain velocity profile takes the form $u_x(z) = u_x(H/2) \frac{\sinh(z/\ell)}{\sinh[H/(2\ell)]}$. The relaxation length $\ell$ is plotted in Fig.~\ref{Fig1}c as a function of $\mathcal{Y}_b$ for frictional and frictionless systems \cite{BTCCA13}. It is observed to  diverge as $\mathcal{Y}_b \to 1$ with an exponent $-1/2$.

This set-up presents similarities with the Couette cell used in \cite{RFP11}, in which the inner cylinder forces a fluid boundary layer while the bulk remains below yield conditions. A (roughly) exponential velocity profile is observed as well.  It should be noted however that the radial profile of the yield parameter $\mathcal{Y}$ is not flat in a Couette cell while $\mathcal{Y}$ is homogeneous in the bulk of our numerical shear cell.

\section{A micro-rheometer}
We have designed a numerical micro-rheometer, similar in spirit to that proposed experimentally in \cite{RFP11}. Bulk horizontal forces acting on two lines of grains, located at a distance $H_b$ away from the boundaries, are added in order to impose a different shear stress in a small central region (Fig.~\ref{Fig2}a). The thickness of this region is denoted as $2H_m$. In Fig.~\ref{Fig2}b, we display the resulting profile of the yield parameter $\mathcal{Y}$ across the cell. It shows transitions from the bulk value $\mathcal{Y}_b$ outside the micro-rheometer to another imposed value, denoted as $\mathcal{Y}_m$, inside the micro-rheometer. We shall emphasize again that, by means of these additional bulk forces, the parameters $\mathcal{Y}_b$, $\mathcal{Y}_m$, $H_b$ and $H_m$ can be arbitrarily chosen and varied. Another important quantity is the value of the shear rate at the locations where $\mathcal{Y}$ crosses $1$. It is inherited from the stress in the boundary layers. We denote it as $\dot \gamma_{b}$ and it is considered as a boundary condition for the bulk of the system. The stress distribution in the cell is associated with velocity and shear rate profiles (Figs.~\ref{Fig2}c and \ref{Fig2}d). From the simulations, we can measure the shear rate $\dot \gamma_{m}$ in the center of the micro-rheometer, and its relation with the rescaled shear stress $\mathcal{Y}_{m}$ gives access to the micro-rheological properties. In practice, $\dot \gamma_{m}$ depends also on $\dot \gamma_{b}$, $\mathcal{Y}_b$, $H_b$ and $H_{m}$.

\begin{figure*}[t!]
\includegraphics{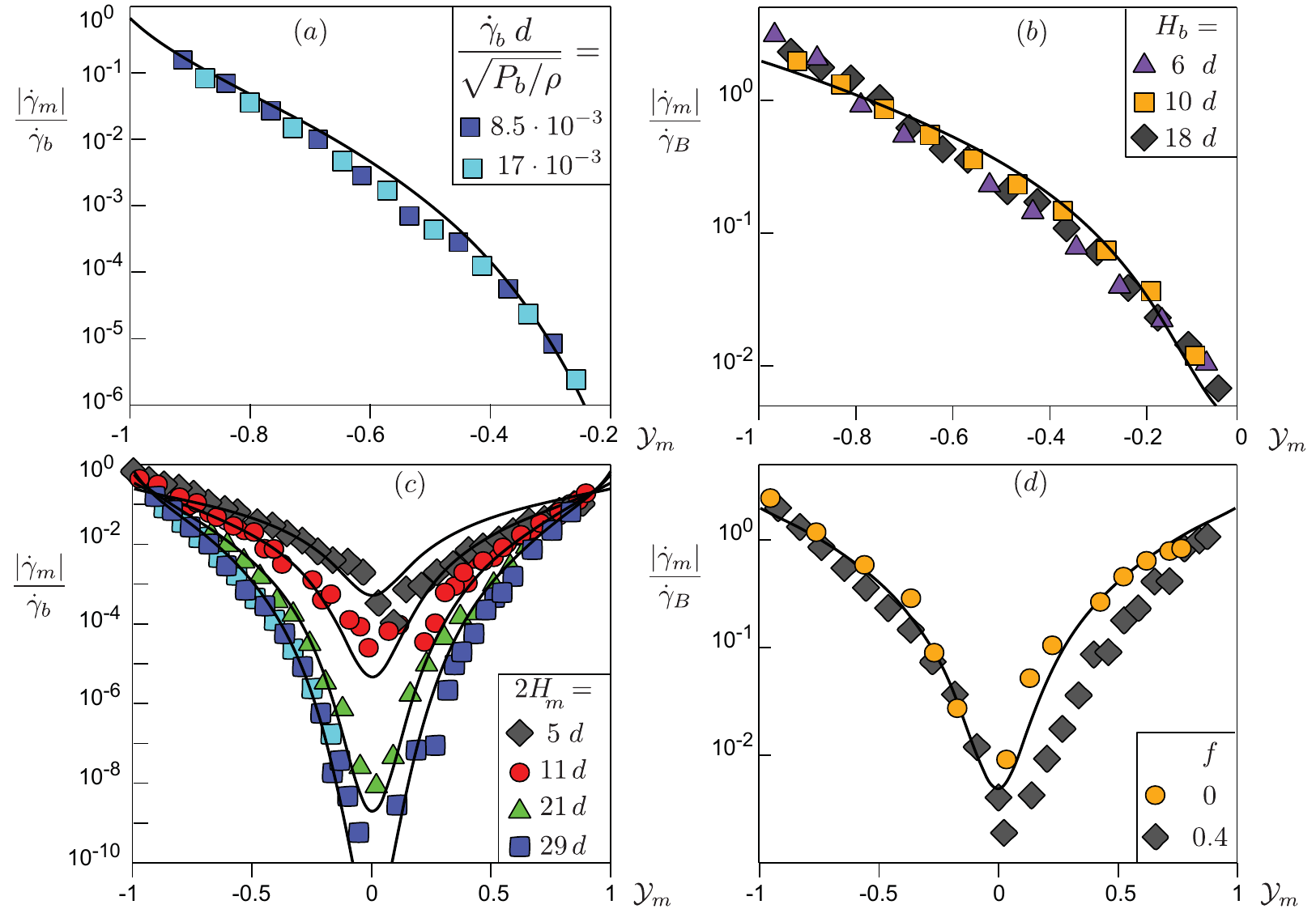}
\vspace{0 mm}
\caption{Shear rate $\dot \gamma_m$ as a function of the yield parameter $\mathcal{Y}_m$ in the micro-rheometer, for $\mathcal{Y}_b=0.9$. (a) Test of the response linearity with respect to the driving shear rate $\dot \gamma_b$; data for $2H_m=29d$ and $H_b=7d$. (b) Effect of the distance $H_b$: data collapse once rescaled by the factor $\dot \gamma_B \equiv \dot \gamma_b\;\exp\left(-H_b/\ell\right)$; data for $2H_m=5d$. (c) Influence of the size $H_m$ of the micro-rheometer; values of $H_m$ in legend and $2(H_b+H_m) \simeq 43 d$. (d) Influence of inter-particle friction coefficient $f$; data for $2H_m=5d$ and $H_b=19d$. In all panels, the solid lines are the predictions of the non-local rheology of \cite{BTCCA13} (Eq.~\ref{gammadotm}), without any adjustable parameter.}
\vspace{0 mm}
\label{Fig3}
\end{figure*}

In the experiments of Reddy~\textit{et al} \cite{RFP11}, the micro-rheometer was a small rod immersed in the grains and submitted to an external force, which, once rescaled by its critical value, plays the role of $\mathcal{Y}_{m}$. The rod creep velocity is the analogous of $\dot \gamma_{m}$ and the shear rate $\dot \gamma_{b}$  is the analogous of the rotation velocity of the inner cylinder. The following key observations reported in \cite{RFP11} are recovered in the numerics. (i) The shear rate $\dot \gamma_{m}$ in the micro-rheometer is proportional to the shear rate $\dot \gamma_{b}$ imposed by the boundary (Fig.~\ref{Fig3}a). (ii)  $\dot \gamma_{m}$ (roughly) decreases exponentially with the distance to the yield condition, measured by $1-|\mathcal{Y}_{m}|$ (Fig.~\ref{Fig3}b). (iii) $\dot \gamma_{m}$ decreases exponentially with $H_b$ (Fig.~\ref{Fig3}b), as evidenced by the data collapse when $\dot \gamma_{m}$ is rescaled by $\dot \gamma_B \equiv \dot \gamma_b\;\exp\left(-H_b/\ell\right)$, using  the relaxation length $\ell$ corresponding to the yield parameter $\mathcal{Y}_b$ (Fig.~\ref{Fig1}c). Note that, for later use, we denote as $\ell_m$ the relaxation length corresponding to $|\mathcal{Y}_m|$. Finally, we find that the curves obtained with frictionless grains are very similar and share these properties (Fig.~\ref{Fig3}d).

\section{Predictions of the non-local rheology}
Can these properties be recovered theoretically using the non-local constitutive relations proposed in \cite{BTCCA13}? The key idea is to extend the rheology $\tau/P=\mu(I)$ by introducing a field $g$ reflecting the local degree of fluidity.  The relative fluidity is defined as $\kappa \equiv d^2 (\nabla^2 g)/g$, which is the lowest order intensive operator.  $\kappa$ is a dimensionless number that remains finite as $g \to 0$, and which reflects the fluidity of the neighborhood of the point considered. Using the parsimony principle, we consider the choice $g=I$ and therefore writes
\begin{equation}
\mathcal{Y} = \frac{\mu(I)}{\mu_c} \left [ 1 - \chi(\kappa) \right ] \quad {\rm with}\quad \kappa \equiv d^2 \frac{\nabla^2 I}{I}
\label{EqNonLocal}
\end{equation}
where $\chi$ is the non-local correction function. At linear order in $\kappa$, we have $\chi(\kappa) \sim \nu \kappa$, where $\nu$ is a phenomenological (positive) constant.

To predict the velocity profile in the bulk and in the micro-rheometer, one needs to linearize the non-local rheology around the static state ($I=0$). The solution of the linearized equations governing the shear rate $\dot \gamma$ is a linear superposition of exponentials of the form $\exp( \pm z/\ell)$, where $\ell$ is the relaxation length displayed in Fig.~\ref{Fig1}. This exponential form is a consequence of linearity and homogeneity and does not depend on any modeling detail. In the micro-rheometer and in the bulk, the solutions respectively read:
\begin{eqnarray}
\dot \gamma & = & \dot \gamma_m \cosh\left( z/\ell_m \right),
\label{gammadotmicrorheometer}\\
\dot \gamma & = & \dot \gamma_+ \exp\left(z/\ell\right)+\dot \gamma_- \exp\left(-z/\ell\right).
\label{gammadotbulk}
\end{eqnarray}
The constants $\dot\gamma_\pm$ are selected by the conditions at the interface $z=H_m$ between the bulk and the micro-rheometer, where the yield parameter presents a discontinuity. In the framework of the model developed in \cite{BTCCA13}, the inertial number $I$ and its derivative $dI/dz$ must be continuous at this interface. One observes in Fig.~\ref{Fig2}d that these profiles nicely fit the numerical data. In particular, as the pressure does not vary at the interface, $|\dot \gamma|$ is correctly predicted to be continuous, although $\dot \gamma$ changes sign. This continuity condition will be discussed at length in the following.

The final expression of the shear rate at the center of the micro-rheometer can be written in a compact form in the limit $H_b\gg\ell$:
\begin{equation}
\frac{{\dot \gamma}_{b}}{|{\dot \gamma}_{m}|} = \frac{1}{2} \,
e^{\frac{H_b}{\ell}}
\left[\cosh\!\left( \frac{H_{m}}{\ell_m}\right) \! +  \frac{\ell}{\ell_m}\sinh\!\left( \frac{H_{m}}{\ell_m}\right) \! \right] \!.
\label{gammadotm}
\end{equation}
This expression is displayed in Fig.~\ref{Fig3} and is in excellent agreement with the numerical data, especially as there is no adjustable parameter once the curve $\ell(\mathcal{Y}_b)$ is calibrated beforehand, see Fig.~\ref{Fig1}c. This analysis predicts the proportionality of ${\dot \gamma}_m$ to ${\dot \gamma}_b$ (property i) as a  consequence of the linearization of the rheological equation. It explicitly predicts that the influence of the distance $H_b$ can be factorized and is exponential (property iii), due to the spatial relaxation of ${\dot \gamma}$ in the bulk. The fast exponential-like decay of ${\dot \gamma}_m$ with  $1-|\mathcal{Y}_m|$ (property ii) also results from the spatial relaxation of the shear rate, however this time, inside the micro-rheometer.

An important consequence of this last point is that the size $H_m$ of the micro-rheometer has a strong influence on $\dot \gamma_{m}$. We have investigated this size effect in our simulations (Fig.~\ref{Fig3}c): the wider the micro-rheometer, the faster the decay of $\dot \gamma_{m}$ with the distance to yield conditions. To understand this property, one needs to disentangle the influence of $H_b$ and $H_m$. Since $\dot \gamma_{m} \simeq \dot \gamma_{b} \exp \left( -H_b/\ell - H_m/\ell_m \right)$, increasing $H_b$ at constant $H_b + H_m$ leads to a decrease of $\dot \gamma_{m}$ if $\ell>\ell_m$. It is the case in Fig.~\ref{Fig3}c as $\ell \simeq 10d$ ($\mathcal{Y}_b=0.9$), while $\ell_m \lesssim 2.5d$ for $|\mathcal{Y}_m|<0.5$.

The size of the probe has also been varied in the experiment reported in \cite{RFP11}: a larger diameter actually facilitates the motion of the probe. In a Couette cell, the stress decreases away from the center: the inner cylinder is just above the yield stress while the bulk of the cell is far below it. Consistently with the measured velocity profile, the relaxation length $\ell$ is small, say $<4d$. When the probe is kept in the vicinity of the critical force  (${\mathcal Y} \lesssim 1$), the relaxation length $\ell_m$ is larger than $\ell$. Smaller micro-rheometers are then slower. As a general conclusion, non-locality is directly reflected by a dependence of the apparent rheological response with the size of the probe. Micro-rheology measurements, where the mobility of intruders in complex fluids is monitored, may then not be a proper rheological tools.

\begin{figure}[t!]
\includegraphics{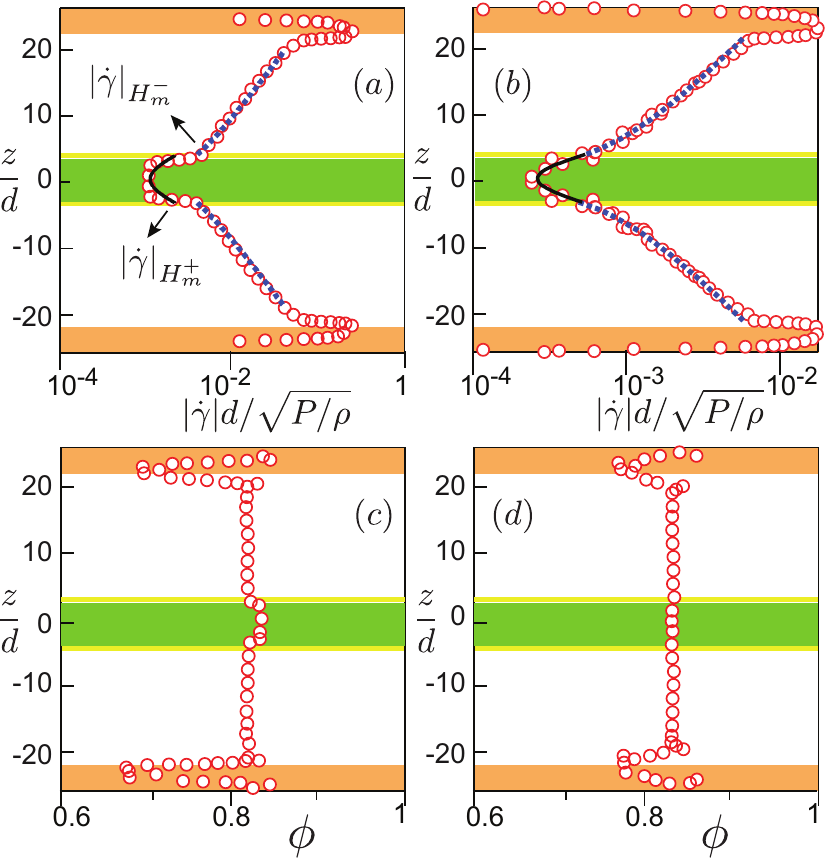}
\vspace{0 mm}
\caption{Shear rate profiles when the grains in the micro-rheometer are sheared (a) in the same ($\mathcal{Y}_m=0.6$) or in the opposite ($\mathcal{Y}_m=-0.5$) direction as the bulk ($\mathcal{Y}_b=0.9$). Symbols: numerical data. Solid lines: prediction of the model inside the micro-rheometer (Eq.~\ref{gammadotmicrorheometer}). Dashed lines: fit of the model in the bulk (Eq.~\ref{gammadotbulk}), leaving $\dot\gamma_\pm$ adjustable. NB: the lengths $\ell = 8.3 d$, $\ell_m = 2.4 d$ (a) and $\ell_m = 3.0d$ (b) are fixed, according to the calibration curve (Fig.~\ref{Fig1}c). Corresponding profiles of the volume fraction for $\mathcal{Y}_m=0.6$ (c) and $\mathcal{Y}_m=-0.5$ (d).}
\vspace{0 mm}
\label{Fig4}
\end{figure}

For frictionless grains ($f=0$), the rheological curves shown are symmetric functions of $\mathcal{Y}_{m}$, as predicted by the model (Fig.~\ref{Fig3}d). However, in the frictional case, the fit of the theory to the data is better when the micro-rheometer is sheared in the direction opposite to that of the bulk ($\mathcal{Y}_m<0$) than for $\mathcal{Y}_m>0$. This discrepancy originates from the continuity condition at the interface of the micro-rheometer. As illustrated in Fig.~\ref{Fig4}, we have used expressions (\ref{gammadotbulk}) and (\ref{gammadotmicrorheometer}) to determine the limit values of $|\dot \gamma|$ on the upper and lower sides of $z=H_m$, denoted below as $|\dot \gamma|_{H_m^\pm}$. One observes that both the \emph{absolute value} of the shear rate $|\dot \gamma|$ and the volume fraction $\phi$ are continuous when $\mathcal{Y}_m<0$ (Fig.~\ref{Fig4}b,d). In this case, one recovers the overall shear rate profile, computed under the assumption of continuity of $I$ and $dI/dz$ at $z=\pm H_m$, which fixes the values of $\dot\gamma_\pm$ (Fig.~\ref{Fig2}d). By contrast, when $\mathcal{Y}_m>0$, both $|\dot \gamma|$ and $\phi$ exhibit a discontinuous jump (Fig.~\ref{Fig4}a,c).

To quantify this discontinuity, we have defined the ratio $\mathcal{R}\equiv |\dot \gamma|_{H_m^+}/|\dot \gamma|_{H_m^-}$. Fig.~\ref{Fig5} shows that $|\dot \gamma|$ is continuous ($\mathcal R=1$) in the frictionless case. $|\dot \gamma|$ is  also continuous in the frictional case, when $\mathcal{Y}_m<0$. When $\mathcal{Y}_m>0$, however, $\mathcal{R}$ is observed to increase exponentially with $\mathcal{Y}_m$. As expected, when the stress discontinuity disappears, i.e. $\mathcal{Y}_m \to \mathcal{Y}_b$, $\mathcal{R}$ tends back to $1$. Interestingly, the anisotropy of the non-local response has been reported in a recent experiment \cite{WH14}.

\section{Discussion}
What can the numerical experiment reported here tell us about the dynamical mechanisms controlling non-locality? First, the exponential behaviors found either experimentally~\cite{RFP11} or in the present simulations are not necessarily Boltzmann-like factors. They result from the linear relaxation of the shear rate, a property shared by all non-local models. As a consequence, these exponentials do not point to a specific mechanism, e.g. mechanical activation. Second, to discriminate between the different models, one needs to highlight the differences in their construction hypotheses and focus on the details of their results. The first fundamental difference is the choice of the order parameter $g$, which characterizes the local degree of fluidity of the system. In the \textit{partial fluidization theory}  \cite{at2001,volfson2003,aranson2008,AT06}, the fluidity $g$ is a phase field appearing in the constitutive relation. In the \textit{fluidity theory} \cite{KK12}, $g$ is defined as the ratio of the shear rate $\dot \gamma$ to the stress ratio $\tau/P$. Both of these approaches are based on a Ginzburg-Landau equation whose control parameter is  $\mathcal Y$. By linearization, they can be expressed under the form:
\begin{equation}
\nabla^2 g = \frac{g}{\ell^2},
\label{EqGene}
\end{equation}
where $\ell$ is a length function of the $\mathcal{Y}$. Solutions invariant along $x$ are exponentials, as found here, so that the differences between the models is not to be found in the shape of the velocity profiles. The partial fluidization theory hypothesizes a subcritical rigidity transition, which means that $\ell$ is expected to diverge at a static threshold $\mathcal{Y}_s>1$, in contrast with our data. Moreover, a subcritical transition would lead to a sharp interface separating a fluid from a solid region, which is not observed. In the fluidity theory, the relaxation length for $\mathcal{Y}<1$ is a characteristic feature of the solid-like state that must be calibrated independently of the liquid-state ($\mathcal{Y}>1$), because the mechanisms controlling the dissipation are different in the two states. It is worth noting that this theory does not account for the hysteresis of the friction coefficient.

By contrast with these models, the \textit{gradient expansion} theory \cite{BTCCA13} emphasizes that the inertial number $I$, the volume fraction $\phi$ or the mean number of contacts $Z$ can be considered as state variables but not the shear stress $\tau$. Using the parsimony principle and arguing that the fluidity must be dimensionless, the order parameter $g=I$ was chosen. The choice $g=\phi_c-\phi$ where $\phi_c$ is the critical volume fraction is completely equivalent as $\phi_c-\phi$ and $I$ are linearly related. The theory is based on a gradient expansion of the rheology with respect to $g$ under two assumptions \cite{BTCCA13}: (i) the mechanical influence of a point on its  neighborhood is roughly isotropic and (ii) the non-local correction remains finite when $g$ vanishes. There is no mechanical difference in the flow above and below the yield conditions. Nothing is said about the mechanical behavior of the solid phase, which can be introduced in the model at a later stage. In particular, the linearization of eq.~\ref{EqNonLocal} is achieved, for $\mathcal{Y}<1$, around a \textit{marginally liquid} state where $g$ tends to $0$ but never vanishes. Contrarily to previous models, the divergence of $\ell(\mathcal{Y})$ as $ d /||\mathcal{Y}|-1|^{1/2}$ is a prediction. Finally, it is fundamental to note that Eq.~(\ref{EqNonLocal}) does not reduce to Eq.~(\ref{EqGene}) when linearized.

\begin{figure}[t!]
\includegraphics{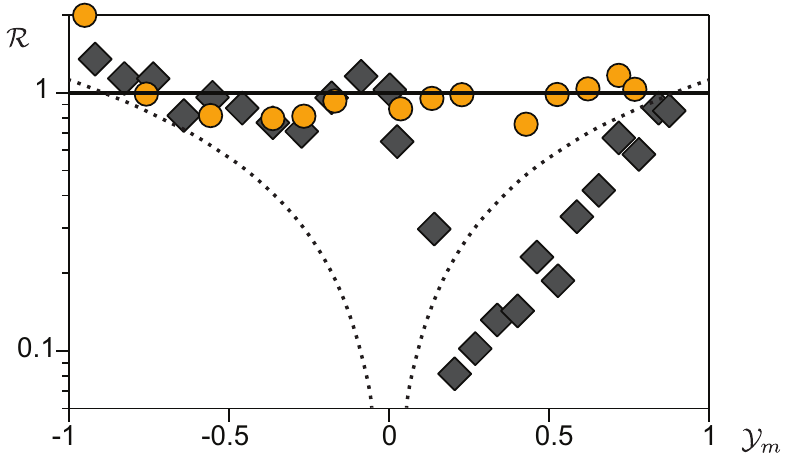}
\vspace{0 mm}
\caption{Ratio $\mathcal{R}$ of the shear rate on the outer and inner sides of the micro-rheometer boundary as a function of the yield parameter $\mathcal{Y}_m$. Symbols and parameters: same as in Fig.~\ref{Fig3}d. Thick solid line $\mathcal{R}=1$: prediction of the gradient expansion model\cite{BTCCA13}. Dotted line: prediction of fluidity theory \cite{KK12}.}
\vspace{0 mm}
\label{Fig5}
\end{figure}

Besides its intrinsic importance, the choice of the fluidity $g$ is reflected by the boundary conditions at a stress discontinuity. Integrating Eq.~(\ref{EqNonLocal})  or Eq.~(\ref{EqGene}) across such a discontinuity, one concludes that $g$ and its gradient must be continuous. In the fluidity theory, $|\dot \gamma|/\mathcal{Y}$ is continuous so that the prediction for our set-up would be: $\mathcal{R}=\mathcal{Y}_m/\mathcal{Y}_b$. The corresponding curve is represented with a dotted line in Fig.~\ref{Fig5} and systematically deviates from observations. The discrepancy is particularly clear around $\mathcal{Y}_m=0$, for which it predicts a static phase in the micro-rheometer whereas our simulations clearly show a flow in these conditions. The fluidity theory does not explain either the shearing asymmetry observed in the frictional case.

Using $g=I$, the gradient expansion theory therefore predicts that $I$ and its derivative are continuous. If the pressure $P$ is constant, the shear rate $|\dot \gamma|$ must be continuous, which imposes that $\mathcal{R}=1$. This prediction is excellent for frictionless grains (Fig.~\ref{Fig5}. For frictional ones, on the other hand, a shear rate discontinuity appears, but only for $\mathcal{Y}_m>0$. This observation can be related to the fact that frictional granular media exhibit a hysteresis between liquid-like and solid-like states. To take this phenomenon into account, one would need to introduce a new state variable, e.g. related to the fraction of sliding contacts,  which can discriminate between a marginal fluid $g \to 0$ and a granular assembly at the onset of avalanching. $g$ would then be a function of $I$ and of this new state variable, allowing for a discontinuity of $I$ (but a continuity of $g$) at a stress discontinuity. The identification of this modified fluidity is the  object of an ongoing theoretical effort and the present micro-rheology measurements  will constitute for all future propositions, an inescapable benchmark test.

\acknowledgments
This work is funded by the ANR JamVibe and a CNES research grant. BA is supported by Institut Universitaire de France.


\end{document}